\documentclass[12pt,a4paper]{article}
\usepackage{color}
\usepackage{bm}
\usepackage{graphicx}  
\usepackage{dcolumn}   
\usepackage{bm}        
\usepackage[english]{babel}
\usepackage{booktabs, multirow, array}
\usepackage[bottom]{footmisc}
\usepackage{dblfnote}
\usepackage{mathtools}
\usepackage{amsmath}
\usepackage{physics}
\usepackage[mathscr]{eucal}
\DFNalwaysdouble 
\usepackage[colorlinks]{hyperref}
\definecolor{LinkColor}{rgb}{0.75, 0, 0}
\definecolor{CiteColor}{rgb}{0, 0.5, 0.5}
\definecolor{UrlColor}{rgb}{0, 0, 0.75}
\hypersetup{linkcolor=LinkColor}
\hypersetup{citecolor=CiteColor}
\hypersetup{urlcolor=UrlColor}
\usepackage{newtxtext}
\usepackage[slantedGreek]{newtxmath}

\title{Hawking temperature of black holes with multiple horizons}
\author{Chiranjeeb Singha\footnote{chiranjeeb.singha@saha.ac.in}$~^{1}$, Pritam Nanda\footnote{pritam.nanda@saha.ac.in}$~^{1, 2}$ and Pabitra Tripathy\footnote{pabitra.tripathy@saha.ac.in}$~^{1,2 }$
\\
$^{1}${\small{Theory Division, Saha Institute of Nuclear Physics. 1/AF Bidhan Nagar,
Kolkata 700064, India}}\\
$^{2}${\small{ Homi Bhabha National Institute,
Training School Complex, Anushaktinagar, Mumbai 400094, India}}}

\begin{document}
\maketitle


\begin{abstract}

There are several well-established methods for computing thermodynamics in single-horizon spacetimes. However, understanding thermodynamics becomes particularly important when dealing with spacetimes with multiple horizons. Multiple horizons raise questions about the existence of a global temperature for such spacetimes. Recent studies highlight the significant role played by the contribution of all the horizons in determining Hawking's temperature. Here we explore the Hawking temperature of a rotating and charged black hole in four spacetime dimensions and a rotating BTZ black hole. We also find that each horizon of those black holes contributes to the Hawking temperature. The effective Hawking temperature for a four-dimensional rotating and charged black hole depends only on its mass. This temperature is the same as the Hawking temperature of a Schwarzschild’s black hole. In contrast, the effective Hawking temperature depends on the black hole's mass and angular momentum for a rotating BTZ hole.

 \end{abstract}

\section{Introduction}

Even after a considerable period since its discovery \cite{Hawking:1975vcx}, Hawking radiation has retained its significance and relevance. Its importance stems from various reasons. Not only has its existence been observed in systems that are far from resembling black holes, but it has also raised several crucial questions pertaining to black holes. The original derivation by Hawking demonstrated that a black hole emits radiation similar to that of a perfect black body, with a temperature directly proportional to the surface gravity of its outer horizon. As the whole calculation of Hawking is too global, attempts are made for a local calculation without the knowledge of the future geometry of spacetime. One such approach is tunneling formalism \cite{Parikh:1999mf}, which considers the creation of particle-antiparticle pairs near or inside the horizon. As one particle tunnels across the horizon, the other escapes to infinity, with the negative energy of the particle falling into the black hole balanced by the positive energy of the escaping particle. Subsequently, the tunneling probability exhibits an exponential fall in energy, giving rise to a temperature compared to the Boltzmann probability distribution. Derivation of tunneling probability involves evaluating the imaginary component of the action for the classically forbidden emission of s-wave particles across the horizon, and the nonzero contribution comes from the pole that occurs at the horizon. The contribution was solely considered from the outer horizon in the original calculation conducted in \cite{Parikh:1999mf}. This leads us to question what would occur if we included contributions from all the physical horizons in a multi-horizon spacetime. \\

Recently, the existence of a global temperature for multi-horizon spacetimes has been proposed \cite{Volovik:2021upi, Volovik:2021iim, Shankaranarayanan:2003ya, Singha:2021dxe, Choudhury:2004ph, CHABAB2021115305, Azarnia:2021vhc, Singha:2022uny}. Contributions from all horizons determine this global temperature. Previous works primarily focused on scalar particle tunneling to compute Hawking's temperature in such spacetimes. Here, we investigate the tunneling of a Dirac particle to determine whether Hawking's temperature depends on the contributions of both horizons.
We consider the tunneling of a Dirac particle from a rotating and charged black hole in four dimensions of spacetime and a rotating BTZ black hole. We also find that a global temperature can indeed exist for these black holes. Interestingly, the global temperature only depends on its mass for a rotating and charged black hole in four spacetime dimensions. It does not depend on its angular momentum and charge. Thus, in four spacetime dimensions, all rotating and charged black holes with the same mass have the same global temperature, regardless of their differing angular momenta and charges. Moreover, we show that the effective temperature is the same as the Hawking temperature of Schwarzschild's black hole \cite{Hawking:1975vcx}.
In a recent study, it has been demonstrated that the effective Hawking temperature for a charged black hole in four dimensions of spacetime, \emph{i.e.,} the Reissner-Nordström black hole, depends only on the black hole's mass \cite{Volovik:2021upi}. It is independent of its charge.  Interestingly, in this scenario, the effective temperature also matches the Hawking temperature of Schwarzschild's black hole.
In contrast, for rotating BTZ black holes, the global temperature depends not only on the black hole's mass but also on its angular momentum.

In this article, we consider the tunneling of a Dirac particle from a rotating and charged black hole in four dimensions of spacetime and a rotating BTZ black hole. Thus, in Sec. \ref{Dirac}, we briefly review the Dirac equation in the curved spacetime. Using the Dirac equation, in Sec. \ref{Kerr-Newman}, we derive the Hawking radiation for a rotating and charged black hole in four spacetime dimensions. Similarly, in Sec. \ref{BTZ}, we derive the Hawking radiation for a rotating BTZ black hole.  We discuss our results in Sec. \ref{conclusion}.

We will set $c = G =\hbar= 1 $ in our calculations.

\section{Dirac particle in a curved background} \label{Dirac}

In this section, we provide a concise overview of the behavior of a Dirac particle in curved spacetime, closely following the framework outlined in references \cite{Birrell:1982ix, Nakahara:2003nw, Parker:2009uva}. Here we also consider a gauge field $A_\mu$ coupled to gravity also with the Dirac field. The Dirac equation in curved spacetime extends the original Dirac equation formulated in flat Minkowski spacetime. Dirac equation in flat space-time governs the dynamics of spinor fields. In Minkowski's spacetime field theory, the spin of a field can be categorized based on how the field's properties change under infinitesimal Lorentz transformations.  We aim to extend these considerations to curved spacetime, which refers to a general Lorentzian manifold $(\mathcal{M},g)$ while maintaining the connection with the Lorentz group locally. This can be accomplished by utilizing the tetrad ($e_\alpha=e_\alpha^\mu\partial_\mu$) and co-tetrad $(w^\alpha=e^\alpha_\mu dx^\mu)$, also known as the vierbein formalism. The fundamental principle of this approach is to establish a system of normal coordinates, denoted as ${e^\alpha_\mu(p)}$, at every point $p$ in spacetime such that when considering a more general coordinate system, the metric tensor becomes more intricate; nevertheless, it remains connected to flat space-time metric $\eta_{\alpha\beta}$ through the following specific relationship, 
\begin{equation}
    g_{\mu\nu}=e^\alpha_\mu e^\beta_\nu\eta_{\alpha\beta}~;\;\;\;\;\;\;\eta_{\alpha\beta}=e^\mu_\alpha e^\nu_\beta g_{\mu\nu}~,
\end{equation}
where $(e^\alpha_\mu, e^\beta_\nu)$ are the vielbein. Index $(\alpha,\beta)$ are related to the local Lorentz frame index, and $(\mu,\nu)$ is related to space-time indices. In a d-dimensional Riemannian manifold, the metric tensor $g_{\mu\nu}$ possesses $d(d+1)/2$ degrees of freedom, whereas the vielbein $e^\mu_\alpha$ has $d^2$ degrees of freedom. Numerous non-coordinate bases yield the same metric, g, with each base being interconnected to others through local orthogonal rotations $ w^\alpha=\Lambda^\alpha_\beta w^\beta$.
This transformation induces a transformation in vielbein as $e^\alpha_\mu=\Lambda^\alpha_\beta  e^\beta_\mu$.
By considering these facts, we can derive the transformation rule for the connection one-form $\omega_\mu^{\alpha\beta}$ from the definition of torsion two forms $(T^\alpha=dw^\alpha+\omega^\alpha_\beta w^\beta)$ as follows,
\begin{equation}
\omega_{\mu\beta}^\alpha=\Lambda^\alpha_\gamma\omega^\gamma_{\mu\delta}\big(\Lambda^{-1}\big)^\delta_\beta+\Lambda^\alpha_\gamma(\partial_{\mu}\Lambda^{-1})^\gamma_\beta~.
\end{equation}
Now as we know, the presence of gamma matrices in the Dirac equation is crucial because they ensure that the equation retains its symmetry under Lorentz transformations. Also, the inclusion of gamma matrices in the Dirac equation is essential for accounting for the phenomenon of spin. These matrices establish a connection between a spinor's different components and a particle's momentum and energy. This relationship between spin and the gamma matrices is a fundamental aspect of quantum field theory; however, when dealing with curved spacetime, we need to construct a modified version of gamma matrices that maintains covariance, and we can achieve this using a normal coordinate system. In curved space-time, we can define gamma matrice as $\gamma^\mu=e^\mu_a\gamma^\alpha$ where $\gamma^\alpha$ is the usual flat space Dirac matrices. In flat space-time, gamma matrices satisfy the following relation,
\begin{equation}
    \{\gamma^\alpha,\gamma^\beta\}=2\eta^{\alpha\beta}\mathbf{I}~,
\end{equation}
where the gamma matrices are, 
\begin{eqnarray}
 \gamma^0 &=& \begin{pmatrix}
i & 0 \\
0 & -i 
\end{pmatrix},\;\ \gamma^1= \begin{pmatrix}
0 & \sigma^3 \\
\sigma^3 & 0 
\end{pmatrix},\;\ \gamma^2= \begin{pmatrix}
0 & \sigma^2 \\
\sigma^2 & 0\end{pmatrix},\nonumber\\ \gamma^3 &=& \begin{pmatrix}
0 & \sigma^1 \\
\sigma^1 & 0 
\end{pmatrix}~.\label{eqn4}
\end{eqnarray}
Now using the definition of $\gamma^\mu$, the above relation can be generalized to a curved space-time $(\mathcal{M},g)$ as,
\begin{equation}
    \{\gamma^\mu,\gamma^\nu\}=2g^{\mu\nu}\mathbf{I}~.
\end{equation}
Under local Lorentz transformation $\Lambda$,  Dirac spinor at a point p $(p \in \mathcal{M})$ transform as, 
\begin{equation}\label{eqn6}
\psi(p)\rightarrow\mathcal{R}(\Lambda)\psi(p),\;\;\;\; \Bar{\psi}(p)\rightarrow\Bar{\psi}(p)\mathcal{R}(\Lambda)^{-1}~,
\end{equation}
where $\bar{\psi}(p)=\psi(p)^\dagger\gamma^0$ and $\mathcal{R}(\Lambda)$ is the spinor representation of Lorentz transformation. In order to formulate an invariant action, we aim to find a covariant derivative that acts as a local Lorentz vector and undergoes spinor-like transformations as,
\begin{equation}
    \mathcal{D}_\alpha\psi(p)=\mathcal{R}(\Lambda)\Lambda_\alpha^\beta\mathcal{D}_\beta\psi(p)~,
\end{equation}
where $\mathcal{D}_\alpha=\nabla_\alpha+i\frac{q}{\hbar}A_\alpha$. Here $A_\alpha=e^\mu_\alpha A_\mu$ is the gauge field. If we identify such a covariant derivative, we can express an invariant Lagrangian as follows,
\begin{equation}\label{eqn8}
    \mathcal{L} = \Bar{\psi} (-i\hbar\gamma.\mathcal{D} +m)\psi~. 
\end{equation}
Now one can check that the quantity $e^\mu_\alpha\partial_\mu\psi(p)$ transform under $\mathcal{R}(\Lambda)$ as follows,
\begin{equation}
\begin{split}
    e^\mu_\alpha\partial_\mu\psi(p)&\rightarrow \Lambda^\alpha_\beta e^\mu_\alpha\partial_\mu \big(\mathcal{R}(\Lambda)\psi(p)\big)\\
    &=\Lambda^\alpha_\beta e^\mu_\alpha\big(\partial_\mu\mathcal{R}(\Lambda)\psi(p)+\mathcal{R}(\Lambda)\partial_\mu\psi(p)\big)~.
    \end{split}
\end{equation}
Here we choose covariant derivative as,
\begin{equation}
    \nabla_\alpha\psi(p)=e^\mu_\alpha\big(\partial_\mu+\Pi_\mu\big)\psi(p)~,
\end{equation}
where $\Pi_\mu$ is the connection necessary to make the derivative covariant. By utilizing equations (\ref{eqn6}) and (\ref{eqn8}),  we can ascertain that $\Pi_\mu$ satisfies the following transformation,
\begin{equation}
    \Pi_\mu=\mathcal{R}(\Lambda)\Pi_\mu\mathcal{R}(\Lambda)^{-1}-\mathcal{R}(\Lambda)^{-1}\partial_\mu\mathcal{R}(\Lambda)~.
\end{equation}
To determine the specific form of $\Pi_\mu$, we examine an infinitesimal local Lorentz transformation given by $\Lambda^\alpha_\beta=\delta^\alpha_\beta+\epsilon^\alpha_\beta$. Under this transformation, the Dirac spinor transforms as,
\begin{equation}
    \psi(p)=\exp\bigg(\frac{i}{2}\epsilon^{\alpha\beta}\Sigma_{\alpha\beta}\bigg)\psi(p)\approx\bigg(1+\frac{i}{2}\epsilon^{\alpha\beta}\Sigma_{\alpha\beta}\bigg)\psi(p)~.
\end{equation}
Here, we define $\Sigma_{\alpha\beta}=\frac{i}{4}[\gamma_\alpha,\gamma_\beta]$ , representing the spinor representation of the Lorentz transformation generators. The quantity $\Sigma_{\alpha\beta}$  satisfies the following Lie algebra,
\begin{equation}\label{eqn13}
    [\Sigma_{\alpha\beta},\Sigma_{\gamma\delta}]=\eta_{\gamma\beta}\Sigma_{\alpha\delta} -\eta_{\gamma\alpha}\Sigma_{\beta\delta} +\eta_{\delta\beta}\Sigma_{\gamma\alpha} -\eta_{\delta\alpha}\Sigma_{\gamma\beta}~. 
\end{equation}
Under the same Lorentz transformation, $\Pi_\mu$ undergoes the following transformation,
\begin{equation}
\begin{split}
    \Pi_\mu&\rightarrow\bigg(1+\frac{i}{2}\epsilon^{\alpha\beta}\Sigma_{\alpha\beta}\bigg)\Pi_\mu\bigg(1-\frac{i}{2}\epsilon^{\gamma\delta}\Sigma_{\gamma\delta}\bigg)\\
    &-\frac{i}{2}\bigg(\partial_\mu\epsilon^{\alpha\beta}\bigg)\Sigma_{\alpha\beta}\bigg(1-\frac{i}{2}\epsilon^{\gamma\delta}\Sigma_{\gamma\delta}\bigg)\\
    &=\Pi_\mu+\frac{i}{2}\epsilon^{\alpha\beta}[\Sigma_{\alpha\beta},\Pi_\mu]-\frac{i}{2}\big(\partial_\mu\epsilon^{\alpha\beta}\big)\Sigma_{\alpha\beta}~.
    \end{split}
\end{equation}
Now considering the transformation of connection one form under infinitesimal Lorentz transformation (infinitesimal version of equation (\ref{eqn4})) and transformation rule of $\Pi_\mu$ along with the equation (\ref{eqn13}), one can show,
\begin{equation}
    \Pi_\mu=\frac{i}{2}\omega_\mu^{\alpha\beta}\Sigma_{\alpha\beta}~.
\end{equation}
Here we arrive at the Lagrangian, which possesses scalar properties under both coordinate transformations and local Lorentz rotations.
\begin{equation}
\begin{split}
    \mathcal{L}&=\Bar{\psi} (-i\hbar\gamma.\mathcal{D} +m)\psi~\\
    &=\Bar{\psi}(p)\big(-i\hbar\gamma^\alpha\nabla_\alpha -\gamma^\alpha qA_\alpha+m\big)\psi(p)\\
    &=\Bar{\psi}(p)\bigg[-i\hbar\gamma^\alpha e^\mu_\alpha\bigg(\partial_\mu+\frac{i}{2}\omega_\mu^{\gamma\delta}\Sigma_{\gamma\delta}+\frac{iq}{\hbar}A_\mu\bigg)+m\bigg]\psi(p)~.
    \end{split}
\end{equation}
Taking the variation of the Lagrangian with respect to the Dirac field, we get the Dirac equation as,
\begin{equation}\label{eqn17}
    \bigg[-i\hbar\gamma^\alpha e^\mu_\alpha\bigg(\partial_\mu+\frac{i}{2}\omega_\mu^{\gamma\delta}\Sigma_{\gamma\delta}+\frac{iq}{\hbar}A_\mu\bigg)+m\bigg]\psi(p)=0~.
\end{equation}

\section{Hawking radiation in Kerr–Newman space time}\label{Kerr-Newman}

Here we start by considering a rotating and charged black hole in four spacetime dimensions. The Kerr–Newman metric is considered for the rotating and charged black hole spacetime in four-dimension. The Kerr–Newman metric in Boyer-Lindquist coordinates is provided as follows \cite{PhysRevD.94.044036, PhysRevD.35.1171, PhysRevD.31.3135, Babar:2020txt},
\begin{equation}\label{kerr-newman}
\begin{split}
   ds^2=-\left(1-\frac{2Mr-Q^2}{\rho^2}\right)dt^2&-\frac{2 (2Mr-Q^2)a\sin^2{\theta}}{\rho^2}dtd\phi\\
    &+\frac{\Sigma}{\rho^2}\sin^2{\theta}d\phi^2+\frac{\rho^2}{\Delta}dr^2+\rho^2d\theta^2~,
    \end{split}
\end{equation}
where,
\begin{equation}
    \begin{split}
    &\rho^2=r^2+a^2\cos{\theta}^2\\
    &\Delta=r^2-2Mr+a^2+Q^2\\
    &\Sigma=(r^2+a^2)^2-a^2\Delta\sin^2{\theta}~.
    \end{split}
\end{equation}
Here $M$ is the mass of the black hole, $a$ is the angular momentum per unit mass, and $Q$ is the charge of the black hole. The spacetime metric (\ref{kerr-newman}) has to coordinate singularity at $r=r_\pm$, defining the horizon of the rotating and charged black hole in four spacetime dimensions, where,
\begin{equation}
    r_{\pm}=M\pm \sqrt{M^2-a^2-Q^2}~.
\end{equation}
The electromagnetic field tensor for Kerr-Newman spacetime is given by,
\begin{equation}
\begin{split}
    F&=\frac{1}{2}F_{\mu\nu}dx^\mu\wedge dx^\nu\\
    &=\frac{Q(r^2-a^2\cos^2\theta)}{\rho^4}dr\wedge(dt-a\sin^2\theta d\phi)\\
    &-\frac{2Qar\sin\theta\cos\theta}{\rho^4}d\theta\wedge(dt-(r^2+a^2) d\phi)~.
    \end{split}
\end{equation}
The vector potential responsible for this field tensor is
\begin{equation}
    A=A_\mu dx^\mu=-\frac{Qr}{\rho^2}(dt-a\sin^2\theta d\phi)~.
\end{equation}
From the above metric (\ref{kerr-newman}), we find out the tetrads in Kerr-Newman spacetime. The four tetrads in this spacetime are given by,
\begin{equation}
\begin{split}
    &e_0^\mu=\bigg({\sqrt{\frac {\Sigma}{\rho^2 \Delta}}}
,0,0,\frac{(2Mr-Q^2)a}{\rho \sqrt{\Delta\Sigma}}\bigg)\\
&e_1^\mu=\bigg(0,\frac{\sqrt{\Delta}}{\rho},0,0\bigg)\\
&e_2^\mu=\bigg(0,0,\frac{1}{\rho},0\bigg)\\
&e_3^\mu=\bigg(0,0,0,\frac{\rho}{\sqrt{\Sigma}\sin\theta}\bigg)~.
\end{split}
\end{equation}
Now we apply the assumption for the spin-up spinor $\psi$ field in the following manner \cite{Li:2008zra, Kerner:2007rr, DiCriscienzo:2008dm},
\begin{equation}\label{24}
    \psi= \begin{pmatrix}
\alpha(t,r,\theta,\phi) \\
0 \\
\beta(t,r,\theta,\phi)\\
0\\ 
\end{pmatrix}e^{\frac{i}{\hbar}\mathcal{I}(t,r,\theta,\phi)}~.
\end{equation}
Please note that we will focus solely on the spin-up scenario since the spin-down case is analogous. To employ the WKB (Wentzel-Kramers-Brillouin) approximation, we can insert the proposed form for a spinor field into the general covariant Dirac equation (equation (\ref{eqn17})). By dividing the equation by the exponential term and disregarding terms involving $\hbar$, we obtain the following set of four equations (for more details, see appendix \ref{I}), 
\begin{equation}\label{eqn25}
\begin{split}
\alpha\bigg\{i(e^t_0\partial_t+e^\phi_0\partial_\phi)\mathcal{I}+ie^t_0qA_t+ie^\phi_0qA_\phi\bigg\}+\beta e^r_1\partial_r\mathcal{I}=0\\
    \beta(ie^\theta_2 \partial_\theta\mathcal{I}+e^\phi_3\partial_\phi\mathcal{I}+qe^\phi_3A_\phi)=0\\
    \alpha e^r_1\partial_r\mathcal{I}-\beta\bigg\{i(e^t_0\partial_t+e^\phi_0\partial_\phi)\mathcal{I}+ie^t_0qA_t+ie^\phi_0qA_\phi\bigg\}=0\\
   \alpha (ie^\theta_2\partial_\theta\mathcal{I}+e^\phi_3\partial_\phi\mathcal{I}+qe^\phi_3A_\phi)=0~.
    \end{split}
\end{equation}

Please be aware that here $\alpha$ and $\beta$ are not constant, their derivatives and the components of spin connections all have a factor of $\hbar$. Hence, in the WKB approximation, these terms can be neglected to the lowest order. Since we only consider the Dirac field outside the event horizon, the above equations always fulfill the $\Delta>0$ condition. The second and fourth equations indicate that a nontrivial solution is only possible when $(\alpha,\;\beta)\neq 0$. Then from the second and fourth equations, we get,
\begin{equation}
    ie^\theta_2\partial_\theta\mathcal{I}+e^\phi_3\partial_\phi\mathcal{I}+qe^\phi_3A_\phi=0~.
\end{equation}
By examining the first and third equations, it becomes evident that these two equations possess a non-trivial solution for $\alpha$ and $\beta$ only when the determinant of the coefficient matrix becomes zero. Subsequently, we can obtain,
\begin{equation}\label{eqn27}
    \bigg( e^t_0\partial_t\mathcal{I}+e^\phi_0\partial_\phi\mathcal{I}+e^t_0qA_t+e^\phi_0qA_\phi\bigg) ^2-  \bigg(e^r_1\partial_r\mathcal{I}\bigg)^2=0~.
\end{equation}
Since the Kerr-Newman spacetime contains two Killing vectors, $(1,0,0,0)$ and $(0,0,0,1)$, we can employ variable separation for $\mathcal{I}$ in the following manner,
\begin{equation}\label{variable_seperation}
    \mathcal{I}(t,r,\theta,\phi)=-\upomega t+\mathcal{J}\phi+\mathcal{R}(r,\theta)~,
\end{equation}
where $\upomega$ and $\mathcal{J}$ are the Dirac particle’s energy and angular momentum. Now by substituting the given expression for $\mathcal{I}(t,r,\theta,\phi)$ into equation (\ref{eqn27}), we can derive the following result,
\begin{equation}
    \bigg( e^t_0\upomega-e^\phi_0\mathcal{J}-e^t_0qA_t-e^\phi_0qA_\phi\bigg) ^2-  \bigg(e^r_1\partial_r\mathcal{R}\bigg)^2=0~.
\end{equation}
Now we solve the above equation for $\theta=\frac{\pi}{2}$ and get,
\begin{equation}
\begin{split}
     \mathcal{R_\pm}&=\pm\int\frac{\big(  e^t_0\upomega-e^\phi_0\mathcal{J}-e^t_0qA_t-e^\phi_0qA_\phi\big)}{e^r_1}dr\\
    &=\pm\int\frac{1}{\Delta}\bigg( \sqrt {\Sigma}(\upomega-qA_t)-\frac{(2Mr-Q^2)a}{\sqrt{\Sigma}}(\mathcal{J}+qA_\phi)\bigg)dr\\
    &=\pm\int\frac{1}{\Delta}\bigg( \sqrt {\Sigma}\bigg(\upomega+q\frac{Q}{r}\bigg)-\frac{(2Mr-Q^2)a}{\sqrt{\Sigma}}\bigg(\mathcal{J}+q\frac{Qa}{r}\bigg)\bigg)~.
\end{split}
\end{equation}
As $\Delta=r^2-2Mr+a^2+Q^2=(r-r_+)(r-r_-)$, the integrand has two poles at the inner and outer horizons. First, we consider the pole $r=r_+$. Then the imaginary part of $\mathcal{R}_\pm$ is given by, 
\begin{equation}
    Im \mathcal{R}_\pm=\pm \pi \left(\frac{r_+^2+a^2}{r_+-r_-}\bigg(\upomega+\frac{q Q}{r_+}\bigg)-\frac{a}{r_+-r_-}\bigg(\mathcal{J}+\frac{qQa}{r_+}\bigg)\right)~.
\end{equation}
Similarly, if we consider the other pole, i.e., pole at $r=r_-$, then the imaginary part of $\mathcal{R}_\pm$ is,
\begin{equation}
    Im \mathcal{\tilde{R}}_\pm =\pm \pi \left(\frac{r_-^2+a^2}{r_--r_+}\bigg(\upomega+\frac{q Q}{r_-}\bigg)-\frac{a}{r_--r_+}\bigg(\mathcal{J}+\frac{qQa}{r_-}\bigg)\right)~.
\end{equation}
Using the Hamilton-Jacobi method of tunneling \cite{MITRA2007240} now, we can calculate the tunneling probability. The probabilities of Dirac particles to cross the outer horizon from inside to outside and from outside to inside are respectively $\mathcal{P}^+_{out}$ and $\mathcal{P}^+_{in}$, where,
\begin{equation}
    \begin{split}
      \mathcal{P}^+_{out}&=exp\left[-\frac{2}{\hbar}Im \mathcal{I}\right]=exp\left[-\frac{2}{\hbar}Im \mathcal{R_+}\right]\\
      \mathcal{P}^+_{in}&=exp\left[-\frac{2}{\hbar}Im \mathcal{I}\right]=exp\left[-\frac{2}{\hbar}Im \mathcal{R_-}\right]~.
    \end{split}
\end{equation}
Similarly, $\mathcal{P}^-_{out}$ and $\mathcal{P}^-_{in}$ are, respectively, probabilities of crossing the inner horizon towards outward and inward. Then we can write,
\begin{equation}
    \begin{split}
      \mathcal{P}^-_{out}&=exp\left[-\frac{2}{\hbar}Im \mathcal{I}\right]=exp\left[-\frac{2}{\hbar}Im \mathcal{\tilde R_+}\right]\\
      \mathcal{P}^-_{in}&=exp\left[-\frac{2}{\hbar}Im \mathcal{I}\right]=exp\left[-\frac{2}{\hbar}Im \mathcal{\tilde R_-}\right]~.
    \end{split}
\end{equation}
The probability that a Dirac particle emits when it is incident on the outer and inner horizon from inside, respectively,
\begin{equation}
    \begin{split}
      \Gamma_1&=exp\left[-\frac{4}{\hbar}Im \mathcal{R_+}\right]\\
      \Gamma_2&=exp\left[-\frac{4}{\hbar}Im \mathcal{\tilde R_+}\right]~.
    \end{split}
\end{equation}
The total probability of particle emission via tunneling from two horizons is given by,
\begin{equation}\label{eqn33}
\begin{split}
    \Gamma=\Gamma_1 \Gamma_2&=exp\left[-\frac{4}{\hbar}\bigg(Im \mathcal{R_+}+Im \mathcal{\tilde R_+}\bigg)\right]\\
    &=exp\left[-\frac{4\pi}{\hbar}\bigg(\omega(r_++r_-)+qQ\bigg)\right]\\
    &=exp\left[-\frac{4\pi (r_++r_-)}{\hbar}\bigg(\omega+\frac{qQ}{(r_++r_-)}\bigg)\right]\\
    &=exp\left[-\frac{8M\pi (\omega-\omega_0)}{\hbar}\right]~,
    \end{split}
\end{equation}
where $\omega_0=-\frac{q Q}{r_++r_-}=q V_{em}$.
This probability function can be compared with Boltzmann distribution, and one can extract the corresponding temperature as, 
\begin{equation}
    T_H=\frac{\hbar}{8 \pi M}~.
\end{equation}
Here, $T_H$ denotes the effective Hawking temperature, considering contributions from both horizons. We observe that this effective temperature depends only on the mass of the black hole. It does not depend on the black hole's charge and angular momentum. Also, we show that the effective temperature is the same as the
Hawking temperature of a Schwarzschild’s black hole \cite{Hawking:1975vcx}.
\section{Hawking radiation from a rotating BTZ black hole}\label{BTZ}

In this section, we consider a rotating black hole in three spacetime dimensions, specifically the rotating BTZ black hole spacetime. The metric describing the rotating BTZ black hole spacetime is given by \cite{SCarlip_1995, Banados:1992wn, Banados:1992gq, Li:2008ws, Dias:2019ery,Martinez:1999qi},
\begin{equation}
\label{btz}
 ds^2=-\mathcal{N}^2dt^2+\frac{1}{\mathcal{N}^2}dr^2+r^2(d\phi+\mathcal{N}^\phi dt)^2~,
\end{equation}
$\mathcal{N}^2$ be defined as $\left(-M+\frac{r^2}{l^2}+\frac{J^2}{4r^2}\right)=\frac{(r^2-r_+^2)(r^2-r_-^2)}{l^2 r^2}$, and $\mathcal{N}^\phi$ as $-\frac{J}{2r^2}$. Here, M represents the mass of the black hole, a dimensionless quantity, while J denotes the angular momentum. Additionally, the cosmological constant $\Lambda$ is related to the AdS radius $l$ as $\Lambda\equiv -(1/l^2)$. The spacetime given by the metric (\ref{btz}) exhibits two coordinate singularities located at $r=r_\pm$, which define the horizons of the rotating BTZ black hole, where,
\begin{equation}\label{BTZ-horizon}
    r_\pm=\sqrt{\frac{Ml^2}{2}\left(1\pm\left[1-\frac{J^2}{M^2 l^2}\right]^\frac{1}{2}\right)}~.
\end{equation}
Here also we compute the tetrads. The tetrads for this spacetime are given by \cite{Li:2008ws},
\begin{equation}
\begin{split}
    &e_0^\mu=\bigg(\frac{1}{\mathcal{N}}
,0,-\frac{\mathcal{N}^\phi}{\mathcal{N}},\bigg)\\
&e_1^\mu=\bigg(0,\mathcal{N},0\bigg)\\
&e_2^\mu=\bigg(0,0,\frac{1}{r}\bigg)~.
\end{split}
\end{equation}
Similar to the four-dimensional spacetime, we get a set of two Dirac equations (for more details, see appendix \ref{II})
\begin{equation}\label{eqn410}
    \begin{split}
    \alpha e^\phi_2\partial_\phi\mathcal{I}+\beta(e^r_1\partial_r\mathcal{I}+e^t_0\partial_t\mathcal{I}+e^\phi_0\partial_\phi\mathcal{I})=0\\
    \alpha(e^r_1\partial_r\mathcal{I}-e^t_0\partial_t\mathcal{I}-e^\phi_0\partial_\phi\mathcal{I})-\beta e^\phi_2\partial_\phi\mathcal{I}=0~.
\end{split}
\end{equation}
It becomes apparent that these two equations have a non-trivial solution for $\alpha$ and $\beta$ only when the determinant of the coefficient matrix equals zero. Consequently, we can derive,
\begin{equation}
    -(e^\phi_2\partial_\phi\mathcal{I})^2-(e^r_1\partial_r\mathcal{I})^2+(e^t_0\partial_t\mathcal{I}+e^\phi_0\partial_\phi\mathcal{I})^2=0~.
\end{equation}
Here, we apply a similar procedure of Kerr–Newman spacetime to calculate the effective temperature. At first, we calculate the $\mathcal{R}$ and get, 
\begin{equation}
\begin{split}
\mathcal{R_\pm}&=\pm\int\frac{\sqrt{\bigg((\upomega+\mathcal{J}\mathcal{N}^\phi)^2-\frac{\mathcal{J}^2\mathcal{N}^2}{r^2}\bigg)}}{\mathcal{N}^2}dr\\
&=\pm\int\frac{l^2r^2\sqrt{\bigg((\upomega-\mathcal{J}\frac{J}{2r^2})^2-\frac{\mathcal{J}^2\mathcal{N}^2}{r^2}\bigg)}}{(r^2-r_+^2)(r^2-r_-^2)}dr~.
    \end{split}
\end{equation}
Then we calculate the imaginary part of $\mathcal{R}_{\pm}$ both the pole at $r=r_{+}$ and $r=r_{-}$ similarly. We now define as  $Im \mathcal{R_+}+Im\mathcal{\tilde R_+}
$ (equation (\ref{eqn33})) $Im \mathcal{R}^{eff}$. Then for rotating BTZ black hole spacetime $Im \mathcal{R}^{eff}$ is given by,
\begin{equation}
\begin{split}
    Im \mathcal{R}^{eff} &=\frac{\pi}{2}\left[\frac{l^2r_+(\upomega-\mathcal{J}\frac{J}{2r_+^2})}{(r_+^2-r_-^2)}+\frac{l^2r_-(\upomega-\mathcal{J}\frac{J}{2r_-^2})}{(r_-^2-r_+^2)}\right]\\
    &=\frac{\pi l^2}{2(r_++r_-)}\bigg(\upomega+\frac{J}{2r_+r_-}\mathcal{J}\bigg)\\
    &=\frac{\pi l^2}{2(r_++r_-)}\bigg(\upomega+\Omega_{eff}\mathcal{J}\bigg)~,
\end{split}
\end{equation}
where $\Omega_{eff}=\frac{J}{2r_+r_-}$ is the effective angular velocity of the two horizons. Using equation (\ref{eqn33}), it is shown that the total probability of particle emission via tunneling from two horizons is given by,
\begin{equation}
    \Gamma =  exp \left[-\frac{2\pi}{\hbar}\frac{l^2}{(r_++r_-)}\bigg(\upomega+\Omega_{Heff}\mathcal{J}\bigg)\right]~.
\end{equation}
This probability function can be likened to the Boltzmann distribution,  allowing one to derive the associated temperature as follows:
\begin{equation}\label{eqn41}
    T_{H}=\frac{\hbar~(r_++r_-)}{2\pi l^2}~.
\end{equation}
Here, $T_H$ represents the effective Hawking temperature, considering contributions from both horizons. From the above equation (\ref{eqn41}) and equation (\ref{BTZ-horizon}), it is clearly shown that this temperature depends on the mass as well as the angular momentum of the black hole.

\section{Conclusion} \label{conclusion}

In this article, we have calculated the tunneling of a Dirac particle from black holes with multiple horizons. This calculation allows us to comment on the Hawking temperature for those black holes. Here, we studied two types of black hole spacetimes: a rotating and charged black hole in four spacetime dimensions, described by the Kerr-Newman metric, and a rotating black hole in three dimensions, described by the rotating BTZ black hole metric.
We have shown that the effective Hawking temperature for the rotating and charged black hole in four spacetime dimensions depends only on the black hole's mass. It is independent of the charge and angular momentum of the black hole. Interestingly, this effective temperature matches with the Hawking temperature of a Schwarzschild's black hole.
On the other hand, the effective Hawking temperature depends on the black hole's mass and angular momentum for the rotating black hole in three spacetime dimensions.

It would be interesting to extend this formalism to higher-dimensional charged and rotating black holes and check whether the effective Hawking’s temperature depends on the angular momentum and charge of the black hole or not. These we leave for the future.
\section*{acknowledgments}
 CS thanks the Saha Institute of Nuclear Physics (SINP) Kolkata for financial support. We thank the reviewer for all the valuable comments and suggestions that helped us to improve the manuscript's quality.
\appendix
\section{Derivation of equation (\ref{eqn25})}\label{I}
Equation (\ref{eqn17}) is the exact Dirac equation in curved space-time. Now to solve the equation we apply the Hamilton-Jacobi method for that we take the limit $\hbar\to 0$ and consider the equation upto $O(\hbar)$. Here, also we consider a mass-less charged particle, so in our case, $m=0$. Now upon substituting the ansatz (\ref{24}) into the equation (\ref{eqn17}), it becomes evident that within an approximation up to $O(\hbar)$, we can neglect the spin coefficient $\omega_\mu^{\alpha\beta}$. So, we start with an approximated  Dirac equation by neglecting the spin coefficient,
\begin{equation}
    -i\hbar \gamma^\alpha e^\mu_\alpha\bigg(\partial_\mu+\frac{iq}{\hbar}A_\mu\bigg)\psi=0~.
    \end{equation}
If we consider only nonzero tetrad, then the above equation reduces to the following
\begin{equation}\label{eqn48}
\begin{split}
     -i\hbar\bigg(\gamma^0 e^t_0\partial_t+\gamma^0 e^\phi_0\partial_\phi+\gamma^1 e^r_1\partial_r+&\gamma^2 e^\theta_2\partial_\theta+\gamma^3 e^\phi_3\partial_\phi+\gamma^0 e^t_0\frac{iq}{\hbar}A_t\\
     &+\gamma^0 e^\phi_0\frac{iq}{\hbar}A_\phi+\gamma^3e^\phi_3\frac{iq}{\hbar}A_\phi\bigg)\psi=0~.
     \end{split}
\end{equation}
Four Gamma matrices are 
\begin{equation}
\begin{split}
    &\gamma^0=\begin{pmatrix}
i & 0 & 0 & 0\\
0 & i & 0 & 0\\
0 & 0 & -i & 0\\
0 & 0 & 0 & -i\\
\end{pmatrix}
\;\;\;
 \gamma^1=\begin{pmatrix}
0 & 0 & 1 & 0\\
0 & 0 & 0 & -1\\
1 & 0 & 0 & 0\\
0 & -1 & 0 & 0\\
\end{pmatrix}\\
& \gamma^2=\begin{pmatrix}
0 & 0 & 0 & -i\\
0 & 0 & i & 0\\
0 & -i & 0 & 0\\
i & 0 & 0 & 0\\
\end{pmatrix}
\;\;  \;\;\; \gamma^3=\begin{pmatrix}
0 & 0 & 0 & 1\\
0 & 0 & 1 & 0\\
0 & 1 & 0 & 0\\
1 & 0 & 0 & 0\\
\end{pmatrix}~.
\end{split}
\end{equation}
By inserting the values of gamma matrices, we can write the equation (\ref{eqn48}) as,
\begin{equation}\label{eqn50}
    -i\hbar\begin{pmatrix}
    A & 0 & B & C\\
    0 & A & D & -B\\
    B & C &-A & 0\\
    D & -B & 0 & -A \\
    \end{pmatrix}
    \begin{pmatrix}
\alpha(t,r,\theta,\phi) \\
0 \\
\beta(t,r,\theta,\phi)\\
0\\ 
\end{pmatrix}e^{\frac{i}{\hbar}\mathcal{I}(t,r,\theta,\phi)}=\begin{pmatrix}
0 \\
0 \\
0\\
0\\ 
\end{pmatrix}~,
\end{equation}
where A, B, and C are
\begin{equation}
\begin{split}
    &A=i\bigg(e^t_0\partial_t+e^t_0\frac{iq}{\hbar}A_t+e^\phi_0\partial_\phi+e^\phi_0\frac{iq}{\hbar}A_\phi\bigg),\\
    &B=e^r_1\partial_r,\\
   & C= -ie^\theta_2\partial_\theta+e^\phi_3\partial_\phi+e^\phi_3\frac{iq}{\hbar}A_\phi~.\\
   & D= ie^\theta_2\partial_\theta+e^\phi_3\partial_\phi+e^\phi_3\frac{iq}{\hbar}A_\phi~
    \end{split}
\end{equation}
Now, using the expression of A, B, C, and D in equation (\ref{eqn50}), we get
\begin{equation}
\begin{pmatrix}
\alpha\bigg\{i(e^t_0\partial_t+e^\phi_0\partial_\phi)\mathcal{I}+ie^t_0qA_t+ie^\phi_0qA_\phi\bigg\} +\beta e^r_1\partial_r\mathcal{I}+o(\hbar)\\
\beta(ie^\theta_2 \partial_\theta\mathcal{I}+e^\phi_3\partial_\phi\mathcal{I}+qe^\phi_3A_\phi)+o(\hbar)\\
\alpha e^r_1\partial_r\mathcal{I}-\beta\bigg\{i(e^t_0\partial_t+e^\phi_0\partial_\phi)\mathcal{I}+ie^t_0qA_t+ie^\phi_0qA_\phi\bigg\}+o(\hbar)\\
\alpha (ie^\theta_2\partial_\theta\mathcal{I}+e^\phi_3\partial_\phi\mathcal{I}+qe^\phi_3A_\phi)+o(\hbar)
\end{pmatrix}e^{\frac{i}{\hbar}\mathcal{I}(t,r,\theta,\phi)}=\begin{pmatrix}
0 \\
0 \\
0\\
0\\ 
\end{pmatrix}~.
\end{equation}
Thus, we arrive at the following four equations,
\begin{equation}
\begin{split}
\alpha\bigg\{i(e^t_0\partial_t+e^\phi_0\partial_\phi)\mathcal{I}+ie^t_0qA_t+ie^\phi_0qA_\phi\bigg\}+\beta e^r_1\partial_r\mathcal{I}=0\\
    \beta(ie^\theta_2 \partial_\theta\mathcal{I}+e^\phi_3\partial_\phi\mathcal{I}+qe^\phi_3A_\phi)=0\\
    \alpha e^r_1\partial_r\mathcal{I}-\beta\bigg\{i(e^t_0\partial_t+e^\phi_0\partial_\phi)\mathcal{I}+ie^t_0qA_t+ie^\phi_0qA_\phi\bigg\}=0\\
   \alpha (ie^\theta_2\partial_\theta\mathcal{I}+e^\phi_3\partial_\phi\mathcal{I}+qe^\phi_3A_\phi)=0~.
    \end{split}
\end{equation}
\section{Derivation of equation (\ref{eqn410})}\label{II}
Here, we apply a similar procedure for writing an approximated Dirac equation for a rotating BTZ black hole. The approximated  Dirac equation for the rotating BTZ black hole spacetime is then given by,
\begin{equation}
\begin{split}\label{eqn56}
    -i\hbar\gamma^\alpha e^\mu_\alpha\partial_\mu\psi=0\\
    \implies -i\hbar\bigg(\gamma^0 e^\mu_0\partial_\mu+\gamma^1 e^\mu_1\partial_\mu +\gamma^2 e^\mu_2\partial_\mu\bigg)\psi=0~
    \end{split}~.
\end{equation}
There are three gamma matrices in three dimensions $\gamma^i =(i\sigma^2,\sigma^1,\sigma^3)$. Where $(\sigma^1,\sigma^2,\sigma^3)$ are the three spin Pauli matrices. Now, considering the nonzero tetrads for BTZ black hole, we can write equation (\ref{eqn56}) as, 
\begin{equation}
\begin{split}
    -i\hbar\begin{pmatrix}
        e^\phi_2\partial_\phi & (e^t_0\partial_t+e^\phi_0\partial_\phi)+e^r_1\partial_r\\
        e^r_1\partial_r-(e^t_0\partial_t+e^\phi_0\partial_\phi) & - e^\phi_2\partial_\phi
        \end{pmatrix}\begin{pmatrix}
            \alpha(t,r,\phi)\\
            \beta(t,r,\phi)\end{pmatrix}e^{\frac{i}{\hbar}\mathcal{I}(t,\theta,\phi)}=\begin{pmatrix}
                0\\
                0
            \end{pmatrix}\\
            \implies\begin{pmatrix}
                \alpha e^\phi_2\partial_\phi\mathcal{I}+\beta(e^r_1\partial_r\mathcal{I}+e^t_0\partial_t\mathcal{I}+e^\phi_0\partial_\phi\mathcal{I})+o(\hbar)\\
                \alpha(e^r_1\partial_r\mathcal{I}-e^t_0\partial_t\mathcal{I}-e^\phi_0\partial_\phi\mathcal{I})-\beta e^\phi_2\partial_\phi\mathcal{I}+o(\hbar)
            \end{pmatrix}e^{\frac{i}{\hbar}\mathcal{I}(t,\theta,\phi)}=\begin{pmatrix}
                0\\
                0
            \end{pmatrix}~.
\end{split}
\end{equation}
So we get a set of two equations as follows,
\begin{equation}
\begin{split}
    \alpha e^\phi_2\partial_\phi\mathcal{I}+\beta(e^r_1\partial_r\mathcal{I}+e^t_0\partial_t\mathcal{I}+e^\phi_0\partial_\phi\mathcal{I})=0\\
    \alpha(e^r_1\partial_r\mathcal{I}-e^t_0\partial_t\mathcal{I}-e^\phi_0\partial_\phi\mathcal{I})-\beta e^\phi_2\partial_\phi\mathcal{I}=0~.
\end{split}
\end{equation}
 
\nocite{*}
\bibliographystyle{ieeetr}
\bibliography{kerr}
\end{document}